\pdfoutput=1

\documentclass[a4paper,11pt]{article}
\pdfoutput=1 

\usepackage{jheppub} 

\usepackage{graphics}
\usepackage{graphicx}
\usepackage{color}
\usepackage{amsmath}
\usepackage{slashed}
\usepackage{epsfig}
\usepackage{hyperref}

\usepackage{caption}
\usepackage{subcaption}

\begin{document}

\title{Constraining portals with displaced Higgs decay searches at the LHC}

\author{Jackson D. Clarke}
\affiliation{ARC Centre of Excellence for Particle Physics at the Terascale, \\ 
School of Physics, University of Melbourne, 3010, Australia.}

\date{\today}

\emailAdd{j.clarke5@pgrad.unimelb.edu.au}

\abstract{
It is very easy to write down models in which long-lived particles
decaying to standard model states are pair-produced via Higgs decays,
resulting in the signature of approximately back-to-back pairs 
of displaced narrow hadronic jets and/or lepton jets at the LHC.
The LHC collaborations have already searched for such signatures with no observed excess.
This paper describes a Monte Carlo method to reinterpret the searches.
The method relies on (ideally multidimensional) efficiency tables, 
thus we implore collaborations to include them in any future work.
Exclusion regions in mixing-mass parameter space are presented 
which constrain portal models.
}


\maketitle

\flushbottom

\section{Introduction}

One of the primary goals of the LHC programme is to study 
the properties of the recently discovered $\approx125$~GeV state, $h$, in great detail \cite{Aad2012tfa,Chatrchyan2012ufa}.
So far the state is consistent with a standard model (SM) Higgs.
Still, plenty of room remains for new physics, particularly in its decays;
$Br(h\to unobserved)\lesssim 20\%$ is still allowed 
for an otherwise SM Higgs
\cite{Belanger2013xza,Cheung2014noa}.
It is therefore sensible to consider spectacular 
Higgs decay channels for which searches might have high sensitivity.
Of interest in this paper is one such channel: pair production of long-lived states which decay 
on detector length scales \cite{Strassler2006ri}.

Long-lived neutral states decaying to SM particles arise naturally in models with approximate symmetries.
There are four mechanisms often encountered in the literature:
an approximate enhanced Poincar\'{e} symmetry \cite{Foot2013hna}, 
which if exact would completely decouple a hidden sector;
an approximate discrete symmetry, which if exact would produce a stable lightest particle;
a low-energy accidental symmetry, with decays only proceeding via a heavy off-shell mediator, and;
small mass-splittings. 
The first mechanism is associated with the usual portal models.

Testable models in which a SM-like Higgs decays to pairs of these long-lived states are easy to build.
The generic signature is an approximately back-to-back pair of displaced 
narrow hadronic jets and/or lepton jets \cite{ArkaniHamed2008qp}
(although this is certainly not the only possibility),
for which null searches have already been performed by the LHC collaborations
\cite{ATLAS2012av,ATLAS-CONF-2014-041,Aad2015asa,Aad2015uaa,  Aad2012kw,Aad2014yea,  Chatrchyan2012jna,CMS2014wda,CMS2014hka,  LHCb2012gja,LHCb2014nma}.
The peculiarity of the signature presents two complementary challenges:
how do collaborations present their results in the most model-independent way possible? 
and how do phenomenologists reinterpret the results in the context of their own models?
This paper describes a simple Monte Carlo method which has implications for both.
The short message is the following: if the relevant efficiency tables are published, then
phenomenologists need only take Monte Carlo events and fold in these efficiencies
to reinterpret searches.
No displaced decays need be simulated since decay probabilities are easily calculated.

The outline of this paper is as follows.
In Section \ref{SecModels} we write down some simple models of interest with an emphasis on portal models.
In Section \ref{SecResults} we describe a method for recasting displaced searches
by way of example, reinterpreting two ATLAS searches.
Exclusion regions in mixing-mass parameter space are presented 
which apply to portal models;
they are explicitly applied to the Higgs and vector portals in Section \ref{SecPortalLims}.
Section \ref{SecConclusion} concludes.

\section{\label{SecModels}Models}

In this section we write down some models of interest which should serve to illustrate
(a) how easy it is to write down a natural model with long-lived states,
(b) how these states can inherit very different couplings to SM particles, and 
(c) how an extended model can introduce non-trivial final states.
Thereupon it should be clear that displaced searches would benefit from a model-independent approach.

\vspace{2pt}
\textit{1. SM plus real singlet scalar: }
With an extra real singlet scalar field $S$, the SM potential becomes
\begin{align}
 V = -\mu^2 \phi^\dagger\phi + \lambda \phi^\dagger\phi \phi^\dagger\phi 
     + \zeta \phi^\dagger\phi  S^2 + \xi \phi^\dagger\phi  S  
      -\mu_s^2 S^2 + \lambda_s S^4 + \alpha S + \beta S^3  \; ,
\end{align}
where $\phi$ is the Higgs field.
The limits $\alpha,\beta,\xi \to 0$ and $\zeta,\xi \to 0$ are technically natural.

Once $\phi$ develops a non-zero vacuum expectation value (VEV),
the cubic term $\xi \phi^\dagger\phi  S$ induces mixing between the SM Higgs doublet state $\phi'_0$ 
and $S$ to produce mass eigenstates
\begin{align}
 \left(
 \begin{array}{c}
  h \\
  s
 \end{array}
 \right)
 =
 \left(
 \begin{array}{cc}
  \cos\rho 	& -\sin\rho \\
  \sin\rho 	& \cos\rho
 \end{array}
 \right)
  \left(
 \begin{array}{c}
  \phi'_0 \\
  S'
 \end{array}
 \right),
\end{align}
where $\rho\approx\sin\rho\ll 1$ is a mixing angle and $h$ is to be identified with the observed SM-like Higgs. 
Then $s$ decays to SM particles through mixing, with a rate $\propto \rho^2$,
leading to displaced decays when $\rho$ is very small.

If $\mu_s^2<0$, then
to first order in $\alpha$, $\beta$, $\xi$, and $\zeta$,
the mixing $\rho$ and the effective $hss$ coupling $\kappa$ are
\begin{align}
\tan\rho = \frac{\xi v}{|m_h^2-m_s^2|}, &&
\kappa = \zeta v ,
\label{EqMixing1}
\end{align}
where $v\approx246$~GeV.

Terms odd in $S$ can be forbidden by demanding some symmetry, for example $Z_2$ ($S\to -S$) or classical scale invariance. 
Then $s=S$ is stable and a dark matter candidate \cite{Silveira1985rk} unless $\mu_s^2>0$
and an effective cubic term is generated by $S$ acquiring 
a non-zero VEV $\langle S \rangle=v_s$. In this case, at tree-level and to leading order in $\zeta$,
\begin{align}
\tan\rho = \frac{\zeta v v_s}{|m_h^2-2m_s^2|} = \zeta v \frac{m_s}{\sqrt{\lambda_s}|m_h^2-2m_s^2|}, &&
\kappa = \zeta v \frac{m_h^2+4m_s^2}{m_h^2-2m_s^2} .
\label{EqMixing2}
\end{align}
This is the usual Higgs portal model \cite{Foot1991bp,Patt2006fw}.

The phenomenology of GeV-scale scalars is detailed in Ref.~\cite{Clarke2013aya}.
We note the significant QCD uncertainties in the widths for $2m_\pi<m_s\lesssim 4$~GeV.
The decay width of the Higgs to light scalars is
\begin{align}
\Gamma(h\to ss) = \frac{\kappa^2}{32\pi m_h}\sqrt{1-\frac{4m_s^2}{m_h^2}} .
\end{align}
Thus for fixed $Br(h\to ss)$ and $m_s$, $\kappa$ is fully determined.\footnote{Note that the requirement $\lambda_s<4\pi$ implies 
$\rho^2 \gtrsim 2\times10^{-9} \left(\frac{m_s}{\text{GeV}}\right)^2\frac{Br(h\to ss)}{0.1}$
in the case of Eq.~\ref{EqMixing2}.
There is no such implication in the Eq.~\ref{EqMixing1} case.}


\vspace{2pt}
\textit{2. Two-Higgs-doublet model (2HDM) plus singlet scalar: }
After symmetry breaking in the CP-conserving 2HDM, one ends up with five Higgs fields,
of which the neutrals are written $h,H,A$ (see Ref.~\cite{Branco2011iw} for details).
In the decoupling limit, $\alpha \to \pi/2-\beta$, $h$ can be associated with the SM-like Higgs.
To avoid tree-level flavour-changing neutral currents, 
a symmetry is usually demanded to ensure that
each of
$(u_R^i, d_R^i, e_R^i)$ couple to only one of the 
Higgs doublets \cite{Glashow1976nt,Paschos1976ay}. 
The four arrangements, known as Type I, Type II, Lepton-specific and Flipped, 
qualitatively alter the couplings of,
in particular, $H$ and $A$ to the SM fermions (see Table 2 of Ref.~\cite{Branco2011iw}).

In the 2HDM with an extra singlet,
if the singlet develops a non-zero VEV
then its real (imaginary) part can mix with $h/H$ ($A$) 
to form a mass eigenstate $s$ ($a$) which could be very light. 
Depending on their admixture of 2HDM Higgs', 
the fermion coupling type, and the values of $\tan\beta$ and $\alpha$, 
$s$ and $a$ could couple dominantly to 
up-types, down-types, or leptons, as demonstrated in Figures 7--10 of Ref.~\cite{Strassler2013fra}. 
Additionally, if these states are mostly singlet then their lifetimes can be long.

\vspace{2pt}
\textit{3. Dark photon: }
A dark photon $\gamma_d$ 
can arise if an extra $U(1)_d$ gauge group is appended to the SM \cite{Holdom1985ag}. 
The SM Lagrangian can be extended by
\begin{align}
 \mathcal{L}_D= -\frac{1}{4} F'_{\mu\nu} F'^{\mu\nu} - \frac{\epsilon}{2\cos\theta_W} F'_{\mu\nu} B^{\mu\nu},
\end{align}
where $F'^{\mu\nu}$ ($B^{\mu\nu}$) is the $U(1)_d$ (hypercharge) field strength,
$\theta_W$ is the Weinberg angle, and $\epsilon$ is the kinetic mixing parameter.
If the $\gamma_d$ is massive, and $m_{\gamma_d}\ll m_Z$, the couplings of $\gamma_d$ to SM particles 
are photon-like and $\propto\epsilon$.\footnote{In a general model this need not be the case, e.g. it is
possible for the extra $U(1)_d$ field to mass mix with the SM $Z$ boson independently of $\epsilon$ 
\cite{Davoudiasl2012ag,Lee2013fda}.}
The lifetime of such a $\gamma_d$ is $\propto \epsilon^{-2}$,
and if $\epsilon$ is small enough the $\gamma_d$ can be long-lived.

A mass term for $\gamma_d$ can be generated through a 
Higgs mechanism \cite{Englert1964et,Higgs1964pj,Guralnik1964eu}. 
The simplest way is to introduce a complex scalar field with a potential 
\begin{align}
 V = -\mu_s^2 |S|^2 + \lambda_s |S|^4 + \zeta |\phi|^2 |S|^2, \label{EqComplexScalarV}
\end{align}
such that $S$ acquires a VEV, leaving a massive state $s$ and giving mass $m_{\gamma_d}\approx g_d v_s$ to the dark photon,
where $g_d$ is the dark gauge coupling.\footnote{This scenario is natural in the limit $\epsilon,\zeta\to 0$ and $\epsilon\to 0$.
Note that an effective $\zeta$ is generated at the level $\sim g_d g \epsilon^2\times$(loop factor) by the kinetic mixing term.}
In this case, the widths of the SM-like Higgs to $\gamma_d\gamma_d$ and $ss$ are $\propto\zeta^2$,
and it is easy to obtain branchings of $\mathcal{O}(10\%)$
when $2m_s$ and/or $2m_{\gamma_d}$ are less than $m_h$ \cite{Gopalakrishna2008dv,Chang2013lfa,Strassler2013fra}.
If $2m_{\gamma_d}<m_s$ then $s$ decays promptly via $s\to \gamma_d\gamma_d$.
Otherwise, it turns out, for $m_s,m_{\gamma_d}\sim$~GeV it is possible that $\epsilon$ and $\zeta$ take values which result in
$\mathcal{O}(10\%)$ Higgs branchings and either $\gamma_d$, or $s$, or both long-lived 
(see Refs.~\cite{Batell2009yf,Curtin2014cca} for some discussion).

\vspace{2pt}
\textit{4. Other models: }
Here we comment on other contexts in which SM-like Higgs decays to long-lived states have arisen in the literature.
We note that many of these models also predict other long-lived state production mechanisms which can be considered independently.

Long-lived right-handed neutrinos can be pair produced 
and decay to fermion trilinears \cite{Graesser2007yj,Graesser2007pc,Cerdeno2013oya,Maiezza2015lza}.
In R-parity violating supersymmetry, the would-be neutralino dark matter candidate can be long-lived, 
decaying to fermion bi- or trilinears 
(see e.g. \cite{Carpenter2006hs,Kaplan2007ap,AristizabalSierra2008ye,Graham2012th,Ghosh2014ida}).
The neutralinos may be pair produced directly or in a cascade
if the spectrum below $m_h$ is sufficiently complex.
Models of WIMP baryogenesis also generically predict 
long-lived particles which decay via fermion trilinears \cite{Cui2014twa}.
In more complex models with a light hidden sector, such as hidden valley models, cascades are common
\cite{ArkaniHamed2008qp,Chan2011aa,Falkowski2010cm,Falkowski2010gv},
producing large multiplicity final states often associated with missing energy.
In a hidden sector with a confining gauge group one expects bound states \cite{Strassler2006im}.
If the confinement scale is $\sim m_h$ then the SM-like Higgs could decay to hidden hadron pairs or hidden glueballs
which then decay back to SM particles via a heavy mediator (or the Higgs portal) 
on collider length scales \cite{Strassler2006ri,Juknevich2009gg}. 
Such phenomenology is typical of recent ``neutral naturalness'' models
\cite{Craig2015pha,Curtin2015fna,Csaki2015fba}.
If the confinement scale is $\ll m_h$ then one expects hidden jets,
which could result in a very large multiplicity of displaced vertices along with missing energy
\cite{Strassler2006qa,Han2007ae,Schwaller2015gea}.

\section{\label{SecResults}Method and results}

In the following section we describe a Monte Carlo method for 
recasting displaced searches by way of example.
The important point to be made is that phenomenologists cannot reliably
simulate the detector response to displaced decays, and are therefore
reliant upon efficiency information provided by the collaborations.\footnote{This point was also made
(and a similar recast method was used) in Ref.~\cite{Cui2014twa}.} 
The recast examples will serve to highlight which efficiency information is of most interest.

\subsection{Displaced hadronic jets}

The ATLAS Collaboration has presented a search for the displaced hadronic decays of pair-produced
long-lived neutral particles in 20.3 fb$^{-1}$ of data collected at $\sqrt{s}=8$~TeV \cite{ATLAS-CONF-2014-041,Aad2015asa}.
They considered pair production via the parton process $gg\to \Phi \to \pi_v\pi_v$, 
where $\Phi$ is a scalar particle 
and $\pi_v$ is a hidden valley pseudoscalar.
The decay of $\pi_v$ is dominated by $b\bar{b}$ for $m_{\pi_v}\gtrsim 10$~GeV 
(the $c\bar{c}$ and $\tau\tau$ decays are subdominant, see their Table 1).
No excess was observed, and limits were placed on the
branching fraction of $\Phi$ as a function of $m_{\pi_v}$ lifetime.
Presently we describe a method to reproduce the result.

Validation samples of $gg\to h \to ss \to (b\bar{b})(b\bar{b})$ events
in $\sqrt{s}=8$~TeV $pp$ collisions were generated using \textsc{Pythia 8.180} \cite{Sjostrand2006za,Sjostrand2007gs}
with the default tune.
We took $m_h=126$~GeV and $m_s=10,25,40$~GeV to match the ATLAS benchmarks.\footnote{The
accuracy of the Monte Carlo for an $s$ of mass $m_s=10$~GeV decaying directly to $b\bar{b}$ is questionable, 
nevertheless it is possible to force \textsc{Pythia} to do the decay, and it appears that this is what was done in the ATLAS analysis.}

The cuts used in the ATLAS analysis are listed in the auxiliary Table~6 of Ref.~\cite{Aad2015asa}.
We recreate them as follows.
The pair produced long-lived particles are required to have\footnote{Selections 
were made using the \textsc{MadAnalysis5} v1.1.10beta \textsc{SampleAnalyzer} framework \cite{Conte2012fm}.}
\begin{align}
 E_T(s_1)>60\text{ GeV}, && E_T(s_2)>40\text{ GeV}, \label{EqETReq}
\end{align}
where $E_T\equiv E\sin\theta$ is a proxy for the measured transverse energy of the resulting jet,
and the $s$ subscript indicates $p_T$-ordering.
This is a fine approximation except for the (non-hadronic) $s\to\tau\tau$ decays with $\sim10\%$ branching.
We also demand $\Delta R (s_1,s_2) > 0.4$
to ensure well-separated jets; this 
makes very little impact on the benchmarks considered by ATLAS, but will matter as $m_s\to m_h/2$.
None of the relevant remaining cuts, such as on isolation and on electromagnetic fraction,
nor the trigger efficiency can be replicated since
no public tool exists to simulate the detector response to displaced decays.\footnote{Though
some attempts have been made \cite{Falkowski2010gv,Liu2015bma}.}
Therefore the remainder of the analysis
necessarily involves the folding in of efficiencies provided by ATLAS.

The CalRatio trigger \cite{Aad2009oea,Aad2013txa}
was used to search for $\pi_v$ decays at or beyond the edge of the electromagnetic calorimeter.
This trigger selects narrow jets with $E_T\gtrsim 35$~GeV, $\log_{10}(E_{H}/E_{EM})>1.2$,
and a lack of tracks in the inner detector.
The trigger efficiency is given as a function of radial (longitudinal) decay position 
for decays in the barrel (endcap) region 
corresponding to the pseudorapidity region $|\eta|<1.5$ ($1.5<|\eta|<2.5$) in Figure 1 of Ref.~\cite{ATLAS-CONF-2014-041}. 
Based on these plots we take the trigger+reconstruction efficiency of the trigger jet
to be non-zero and constant only between 2.0 to 3.5~m in the barrel and 4.0 to 5.5~m in the endcap, 
with a respective ratio of $0.20/0.06$.
The reconstruction efficiency for the non-trigger jet is not given, but we take it similarly.
By construction, the following quantity is then \textit{proportional} to the trigger/reconstruction probability
for a given $s$ of lifetime $c\tau$:
\begin{align}
 \hat{\varepsilon}(s,c\tau) =
 \begin{cases}
  0.20 P(s,c\tau) \text{  if in barrel,}\\
  0.06 P(s,c\tau) \text{  if in endcap,}
 \end{cases}
\end{align}
where $P$ is the probability that a state $x$ decays between $L_{min}$ and $L_{max}$,
\begin{align}
 P(x,c\tau)=-\exp\left( -\frac{L_{max}}{\gamma\beta c\tau} \right) +\exp\left( -\frac{L_{min}}{\gamma\beta c\tau} \right),
 \label{EqPDecay}
\end{align}
with $\gamma$ and $\beta$ the relativistic parameters for $x$.
The timing of the $s$ decay is required to satisfy $\Delta t<5$~ns with respect to a $\beta=1$ particle. 
This corresponds to requiring an absolute decay distance
\begin{align}
 L_{abs}<\frac{\beta}{1-\beta}1.5\text{ m} \equiv L_{abs}^{max}.
\end{align}
Thus after the aforementioned selection cuts, each event is weighted by a factor
\begin{align}
 W(c\tau) = \hat{\varepsilon}(s_1,c\tau) \hat{\varepsilon}(s_2,c\tau),
\end{align}
where we take
\begin{align}
\left(L_{min},L_{max}\right) &= 
\begin{cases}
\left(
\min\left(
  \frac{2.0\text{ m}}{\sin\theta},
  L_{abs}^{max}
  \right)
,
\min\left(
  \frac{3.5\text{ m}}{\sin\theta},
  L_{abs}^{max}
  \right)
\right) & \text{if in barrel,}\\
\left(
\min\left(
  \frac{4.2\text{ m}}{\cos\theta},
  L_{abs}^{max}
  \right)
,
\min\left(
  \frac{5.2\text{ m}}{\cos\theta},
  L_{abs}^{max}
  \right)
\right) & \text{if in endcap,}
\end{cases}
\end{align}
for each $s$, where $\theta$ is the polar angle from the beam line.
After this, the remainder of the cuts used in the ATLAS analysis 
should be largely independent of $c\tau$.
As such, penultimately, we rescale the events (with a common number for all $c\tau$ and $m_s$) to fit ATLAS results;
the factor turns out to be $\approx 6\times19.0$~pb~$\times20.3$~fb$^{-1}/N_{sim}$, 
where $N_{sim}$ is the number of simulated events.
We ignore for simplicity the additional $m_s$-dependent 
$\lesssim 10\%$ effect related to sub-dominant but non-zero $s\to\tau\tau$ branching.

After these requirements we find good agreement for the $m_s=10,25$~GeV samples, 
but we overpredict for $m_s=40$~GeV. 
This is because as $m_s$ approaches $E_s$ (Eq.~\ref{EqETReq}), the $s$ decay products spread out 
and the narrow jet trigger efficiency decreases.
To properly take this into account we require information 
on how the efficiency depends on the $b\bar{b}$ opening angle,
or equivalently, in the limit $E_s^2 \gg m_b^2$, the boost.
In the absence of such information,
and in an attempt to capture the physics involved,
we demand the following (admittedly crude) 
bound on the opening angle of the $b\bar{b}$ pair from the leading $s$:\footnote{Another 
option is to reject leading $s$ with boost lower than $\gamma_{cut}$, 
forbidding particles which have the potential to 
produce opening angles $\approx \arccos\left[1 - 2/\gamma_{cut}^2 + 8 m_x^2 (\gamma_{cut}^2-1)/(m_s^2 \gamma_{cut}^4)\right]$,
where $m_x$ is the decay product mass.}
\begin{align}
 \Delta R (b,\bar{b}) < 1.5 .
\end{align}
This cut has been tuned so that our results for $m_s=40$~GeV best agree with those of ATLAS.
Note that the spatial separation of the $b\bar{b}$ pair as seen by the hadronic calorimeter
is smaller than such a large $\Delta R$ would normally suggest, since the pair appears late.

\begin{figure}[t]
 \centering
 \begin{subfigure}[t]{.45\textwidth}
  \includegraphics[trim={0 2cm 0 0.5cm},clip,width=\textwidth]{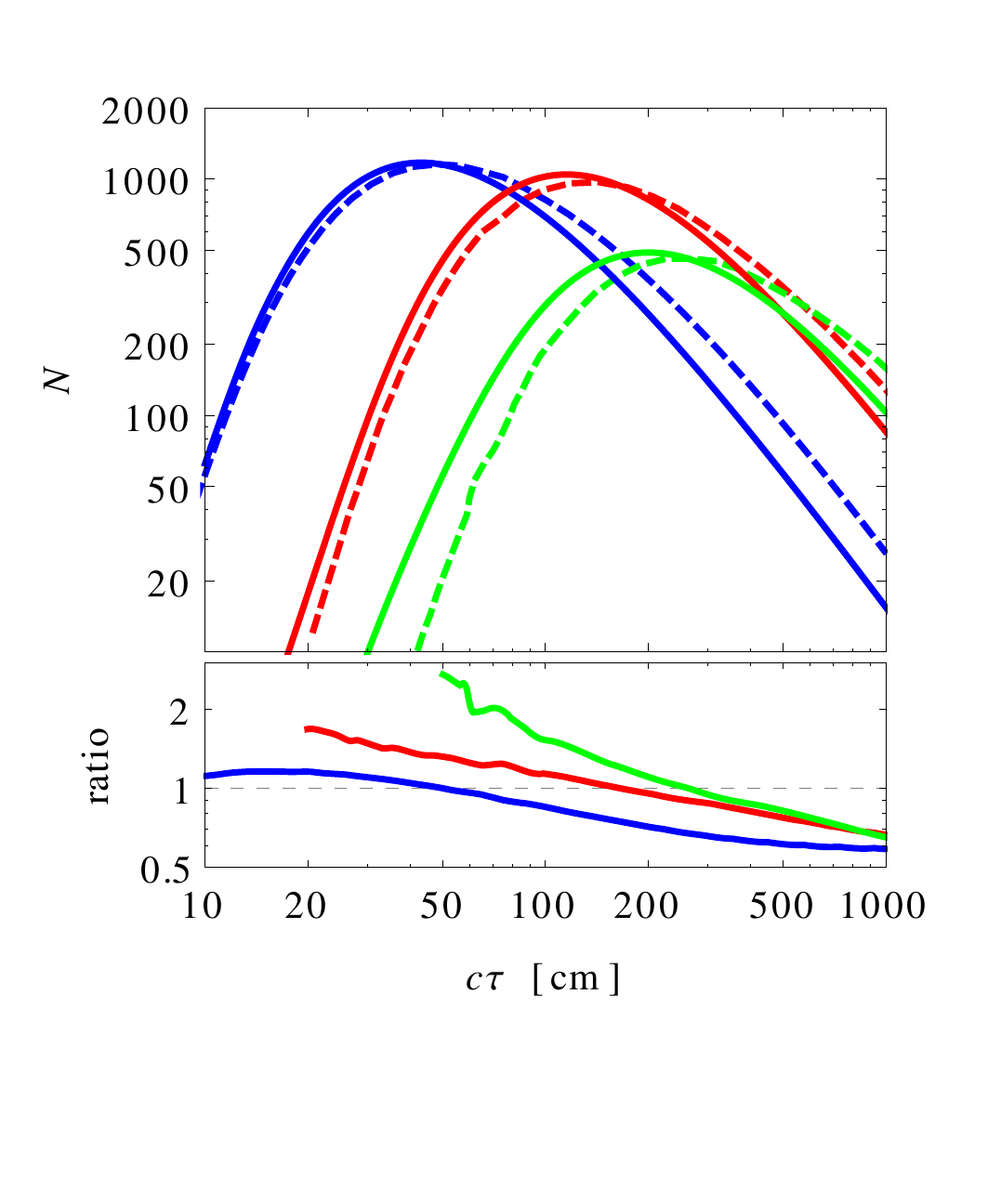}
  \caption{}
  \label{FigHadJets}
 \end{subfigure}
 \begin{subfigure}[t]{.49\textwidth}
  \includegraphics[width=\textwidth]{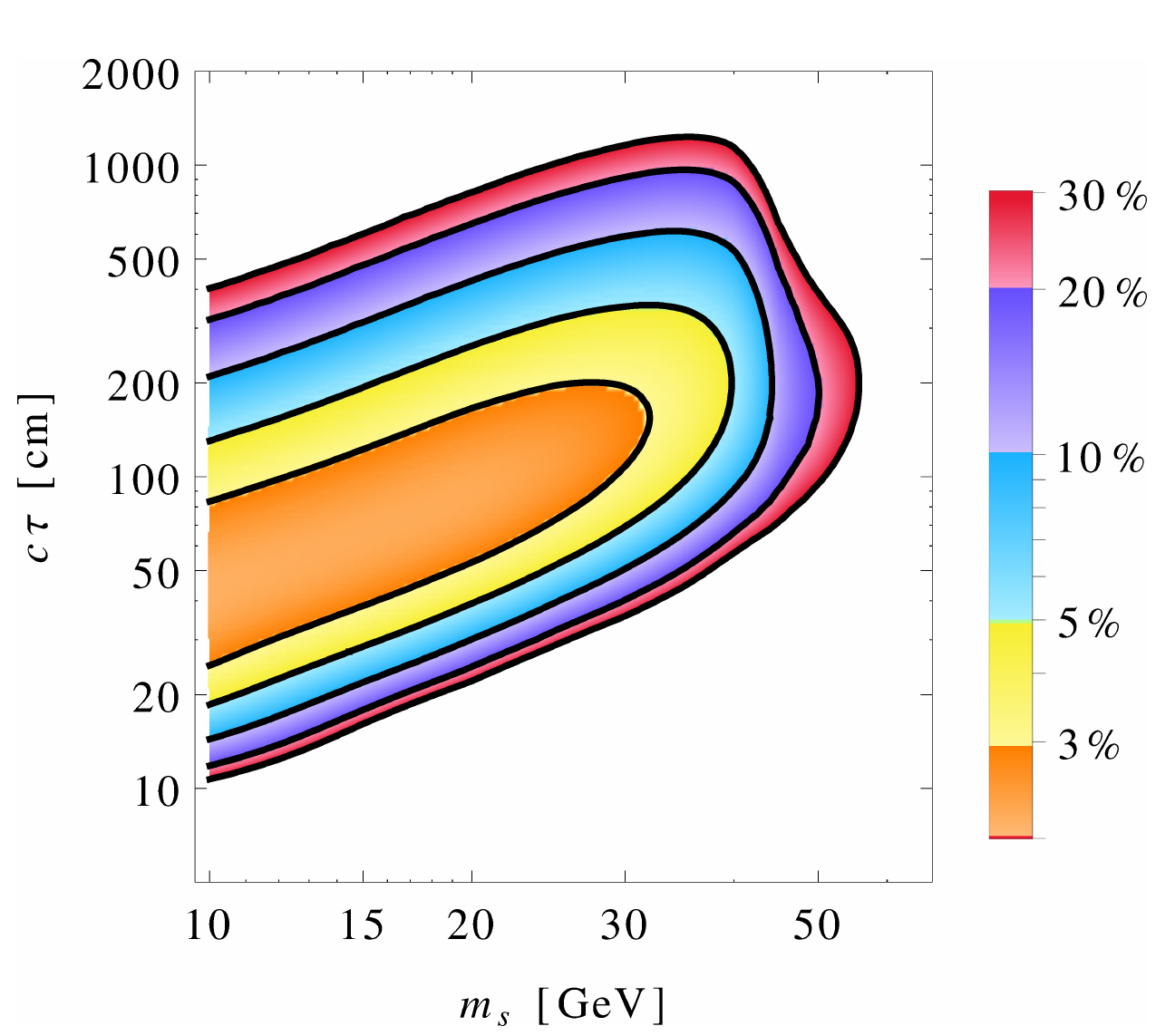}
  \caption{}
  \label{FigHadResults}
 \end{subfigure}
 \caption{(a) Predicted number of events for the displaced hadronic jet analysis assuming 100\% Higgs branching. 
 Upper: the results from ATLAS (dashed) and our results (solid) for $m_s=10,25,40$~GeV beginning left-to-right.
 Lower: the ratio of our results to those of ATLAS.
 (b) Excluded parameter region for $s$ assuming fixed $Br(h\to ss)\times Br(s \to hadronic)^2$; 
 the contours mark branchings of 30\%, 20\%, 10\%, 5\%, 3\%.}
\end{figure}

In Fig.~\ref{FigHadJets}  we compare the number of events predicted by our analysis to those of ATLAS 
assuming 100\% Higgs branching. 
Despite the apparent crudeness of some of our assumptions, 
we observe good agreement.
The 95\% CL limit of 20 events (inferred from the ATLAS plots) can be used to
obtain a limit on the Higgs exotic branching fraction as a function of $c\tau$.
To obtain the exclusion for alternative masses,
\textsc{Pythia} signal samples $gg\to h \to ss \to (b\bar{b})(b\bar{b})$ 
of varying $m_s$ were fed through our analysis.
In Fig.~\ref{FigHadResults} we present our results as limits on 
$Br(h\to ss)\times Br(s\to hadronic)^2$ as a function of $m_s$ and $c\tau$.
This is also a good approximation for the limit on $Br(h\to \gamma_d\gamma_d)\times Br(\gamma_d\to hadronic)^2$.
Given the good match to ATLAS, we are confident that our results for 10~GeV~$<m_s<40$~GeV are reliable, 
and for $m_s>40$~GeV are at least sensible.
In Sec.~\ref{SecPortalLims}, Fig.~\ref{FigHadResults} is reinterpreted to bound mixing-mass parameter space 
for the Higgs and vector portals.

We were fortunate in this analysis because the topology of interest was essentially
already considered by ATLAS for three benchmark values of $m_s$. 
This allowed us to demonstrate the not obvious fact that much of the $c\tau$ dependence
is taken into account simply by reweighting events with easily calculated decay probabilities (Eq.~\ref{EqPDecay}).
In the region where $\gamma_s\sim 1$ we saw that there was an additional
effect that had to be considered,
related to the boost-dependence of the trigger/reconstruction efficiency.
This could have been anticipated, since any momentum dependence was already integrated out of the 
efficiency plots provided by ATLAS.
For this reason our analysis as it stands cannot be reliably reapplied to another model
since the overall efficiency will scale non-trivially with the 
(correlated) $p_T$ distributions of the two long-lived particles.
However it should serve as a conservative estimate for models with more boosted (on average) long-lived pairs.
In the next analysis the $p_T$ dependence of the efficiencies is provided and taken into account.

\subsection{Displaced lepton jets}

In Ref.~\cite{Aad2014yea}, the ATLAS Collaboration presented the search for a SM-like Higgs decaying to a
long-lived pair of $\mathcal{O}(100$~MeV) dark photons
in 20.3 fb$^{-1}$ of data collected at $\sqrt{s}=8$~TeV.
The benchmark process considered was
\begin{align}
gg \to h \to f_{d_1}\overline{f}_{d_1} \to (f_{d_2}\gamma_d)(\overline{f}_{d_2}\gamma_d), \label{EqFRVZprocess}
\end{align}
where the $f_{d_i}$ are hidden fermions and $\gamma_d$ is the long-lived dark photon,
inspired by Falkowski--Ruderman--Volansky--Zupan (FRVZ) models \cite{Falkowski2010cm,Falkowski2010gv}.
Each dark photon decays at or beyond the outer edge of the electromagnetic calorimeter
to either $\mu\mu$ or $ee$/$\pi\pi$,
resulting in a muon jet (Type 0) or a narrow jet (Type 2) respectively.\footnote{Note
that electrons in the hadronic calorimeter will resemble a narrow hadronic jet.}
No excess over the SM expectation was observed and limits were placed on the exotic branching fraction of the Higgs 
as a function of $\gamma_d$ lifetime.
These limits are clearly model-dependent, and it is not clear how the phenomenologist might translate them.
We describe a simple method below.

A validation sample of FRVZ events was generated 
in \textsc{Pythia 8.180} \cite{Sjostrand2006za,Sjostrand2007gs}
by changing the properties of in-built particles.
We took $(m_h,m_{f_{d_1}},m_{f_{d_2}},m_{\gamma_d})=(125,5,2,0.4)$~GeV
and $Br(\gamma_d\to \text{Type 0})=0.45$
to match the ATLAS benchmark.

The selection criteria (cuts) for the ATLAS analysis are detailed in Table 1 of Ref.~\cite{Aad2014yea}.
We recreate the analysis as follows. 
The dark photons are required to satisfy
\begin{align}
 |\eta(\gamma_d)|<2.5 , && |\Delta \phi (\gamma_{d_1},\gamma_{d_2})|>1 ,
\end{align}
as a proxy for the jet acceptance.
The remainder of the analysis necessarily involves the folding in of efficiencies.
After each $\gamma_d$ decays, the final states are of Type 0-0, 0-2, or 2-2 in the obvious way.
After selection, each event is weighted by a factor
\begin{align}
 W(c\tau)=\frac{19.2\text{ pb}\times20.3\text{ fb}^{-1}}{N_{sim}}
 P_{reco}(\gamma_{d_1},c\tau)P_{reco}(\gamma_{d_2},c\tau)\varepsilon_{trig}(\gamma_{d_1},\gamma_{d_2}), \label{eqweightlepjets}
\end{align}
where $P_{reco}(\gamma_d,c\tau)$ is the reconstruction probability\footnote{This term
includes additional rejection criteria such as inner detector isolation.}
for a $\gamma_d$ of lifetime $c\tau$, 
and $\varepsilon_{trig}(\gamma_{d_1},\gamma_{d_2})$ is the trigger efficiency given that the event is reconstructed.
Eq.~\ref{eqweightlepjets} assumes that the reconstruction probability for each of
the lepton jets can be considered independently.
Both $P_{reco}$ and $\varepsilon_{trig}$ depend on the event Type, and will be described presently.

The reconstruction efficiencies for a $\gamma_d$ with transverse momentum $p_T$ decaying at a length $L$,
$\varepsilon_{reco}\equiv \varepsilon_{reco}(p_T,L)$, 
are provided in the ATLAS auxiliary Tables 1--4 \cite{Aad2014yea} for Type 0 and Type 2 jets
decaying in the barrel and endcap regions, as defined in Table~\ref{Tabeffrec}.
We assume $\varepsilon_{reco}=0$ outside of those $L, \eta, p_T$ regions,
which appears to be stricter (and therefore more conservative) 
than the barrel/endcap regions used in the full analysis.
Since the $\gamma_d$ are very boosted, we do not require a timing veto.
The reconstruction probability for each jet is then
\begin{align}
 P_{reco}(\gamma_d,c\tau) = \sum_{\text{$L$ bins}} P(\gamma_d,c\tau) \varepsilon_{reco}(p_T^{\gamma_d},L),
\end{align}
where $P(\gamma_d,c\tau)$ is given by Eq.~\ref{EqPDecay} with
\begin{align}
\left(L_{min},L_{max}\right) &= 
\begin{cases}
\left(
  \frac{L_{xy\text{-bin}}^{min}}{\sin\theta},
  \frac{L_{xy\text{-bin}}^{max}}{\sin\theta}
\right) & \text{if in barrel,}\\
\left(
  \frac{L_{z\text{-bin}}^{min}}{\cos\theta},
  \frac{L_{z\text{-bin}}^{max}}{\cos\theta}
\right) & \text{if in endcap.}
\end{cases}
\end{align}

For any event involving a Type 2 jet, ATLAS used the previously described CalRatio trigger. 
ATLAS provides the CalRatio trigger efficiency $\varepsilon_{cal}$,
defined as the fraction of jets passing the offline selection which also pass the trigger,
separately as a function of $p_T$ and $\eta$.
Type 0-0 events are collected by the 3mu6\_MSonly trigger \cite{Aad2012kw,Aad2014yea},
which requires at least three standalone (not combined with an inner detector track) muons with $p_T>6$~GeV.
The efficiency of this trigger is dominated by the granularity of the muon spectrometer;
to reconstruct three muons at least one of the dark photons must produce a pair of muons which have a discernible opening angle.
ATLAS provides the efficiency $\varepsilon_{2}$, defined as the fraction of $\gamma_d$ passing the offline selection
and also producing two distinguishable muons,
separately as a function of $p_T$ and $\eta$.
The efficiency for detecting at least one muon is quoted as $\varepsilon_{\ge 1}=0.8$ (0.9) in the barrel (endcap) region.
For our purposes we converted these efficiencies, making the assumption of independence, into functions of two variables,
$\varepsilon_{cal}\equiv\varepsilon_{cal}(p_T,\eta)$ and $\varepsilon_2\equiv\varepsilon_2(p_T,\eta)$.
For each event, the trigger efficiency given event reconstruction is taken as
\begin{align}
 \varepsilon_{trig}(\gamma_{d_1},\gamma_{d_2})=
 \begin{cases}
  \varepsilon_{\ge1}(\gamma_{d1})\varepsilon_{2}(\gamma_{d2})
  + \varepsilon_{\ge1}(\gamma_{d2})\varepsilon_{2}(\gamma_{d1})
  - \varepsilon_{2}(\gamma_{d1}) \varepsilon_{2}(\gamma_{d2})
  & \text{ if Type 0-0,} \\
  \varepsilon_{cal}(\gamma_{d_{\text{Type-0}}}) 
  & \text{ if Type 0-2,} \\
  \varepsilon_{cal}(\gamma_{d_1})+\varepsilon_{cal}(\gamma_{d_2})
  -\varepsilon_{cal}(\gamma_{d_1})\varepsilon_{cal}(\gamma_{d_2})
  & \text{ if Type 2-2,} \;
 \end{cases}
\end{align}
where, in an obvious notation, $\varepsilon(\gamma_d) \equiv \varepsilon(p_T^{\gamma_d},\eta^{\gamma_d})$.
This is not quite a model-independent trigger efficiency, 
since $\varepsilon_{cal}$ and $\varepsilon_2$ are derived from a lepton-jet gun event sample,
in which the $\gamma_d$ are generated uniformly in ($p_T,\eta$),
but it serves as a good approximation for our purposes.
After weighting by reconstruction probabilities,
we find that this trigger efficiency for the FRVZ sample 
rescales the number of events by an approximately global number,
$\approx 0.5$ for $c\tau_{\gamma_d} = 0.1$~cm and $\approx 0.3$ for $c\tau_{\gamma_d} = 100$~cm.

\begin{table}
 \centering
 \begin{tabular}{|c|c|c|c|}
  \hline
  \multicolumn{2}{|c|}{Type 0} & \multicolumn{2}{|c|}{Type 2} \\ \hline
  Barrel & Endcap & Barrel & Endcap \\ \hline
  $14\le L_{xy}$/cm~$\le 780$ & $50\le L_z$/cm~$\le 1400$ & $150\le L_{xy}$/cm~$\le 410$ & $350\le L_{z}$/cm~$\le 650$ \\
  $|\eta|<0.9$ & $1.2 < |\eta| <2.5$ & $|\eta|<1.0$ & $1.5<|\eta|<2.4$ \\
  $10\le p_T/\text{GeV}\le 100$ & $10\le p_T/\text{GeV}\le 100$ & $20\le p_T/\text{GeV}\le 100$ & $20\le p_T/\text{GeV}\le 100$ \\ \hline
 \end{tabular}
 \caption{Definitions of barrel and endcap regions for $\varepsilon_{reco}(p_T,L)$ as defined by ATLAS.}
 \label{Tabeffrec}
\end{table}

\begin{figure}[t]
 \centering
 \begin{subfigure}[t]{.44\textwidth}
  \includegraphics[trim={0 2cm 0 0.5cm},clip,width=\textwidth]{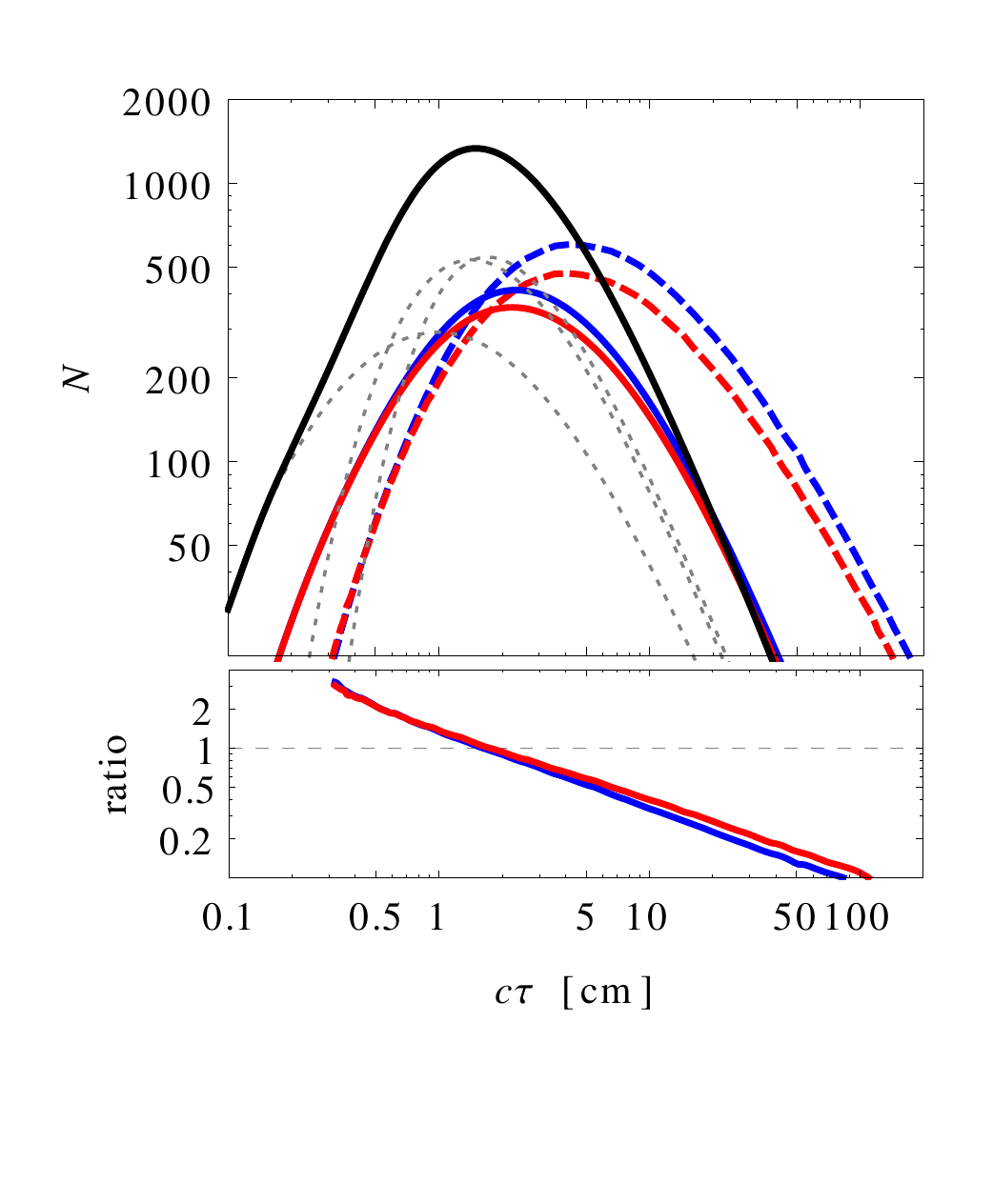}
  \caption{}
  \label{FigLepJets}
 \end{subfigure}
 \begin{subfigure}[t]{.49\textwidth}
  \includegraphics[width=\textwidth]{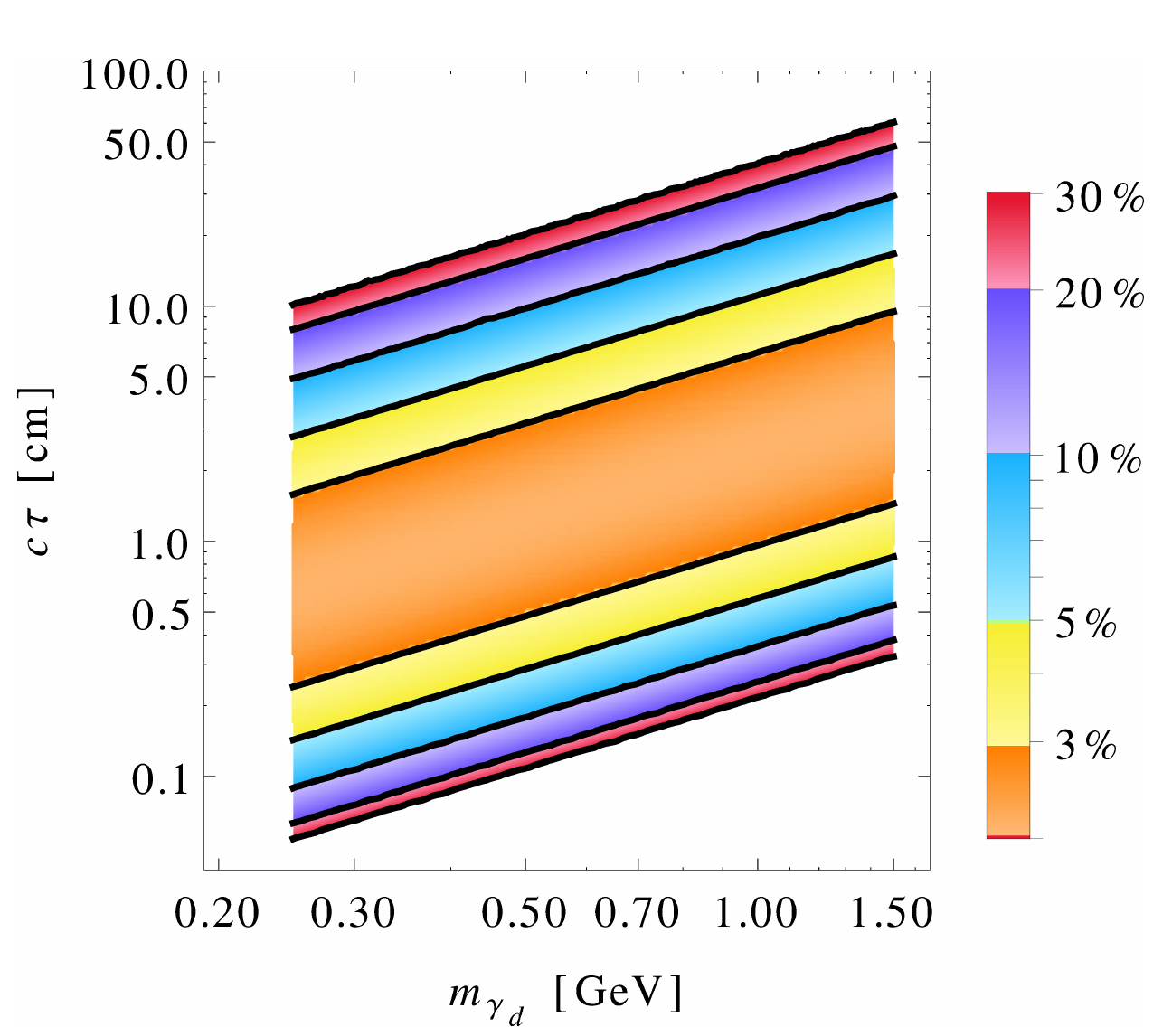}
  \caption{}
  \label{FigLepResults}
 \end{subfigure}
 \caption{(a) Predicted number of events in the lepton jet analysis assuming 100\% Higgs branching,
 $m_{\gamma_d}=400$~MeV, and $Br(\gamma_d\to \text{Type 0})=0.45$. 
 Upper: the FRVZ model results from ATLAS (dashed) and our results (solid)
 for all event Types (blue/upper) and excluding Type 2-2 events (red/lower).
 Also shown are the number of events expected for the $h\to\gamma_d\gamma_d$ model in solid black;
 dotted curves beginning left-to-right show the break down in terms of Type 0-0, 0-2, 2-2 events respectively.
 Lower: the ratio of our FRVZ model results to those of ATLAS.
 (b) Excluded parameter region for $\gamma_d$ assuming fixed $Br(h\to \gamma_d\gamma_d)\times Br(\gamma_d \to \mu\mu)^2$. 
 The contours mark branchings of 30\%, 20\%, 10\%, 5\%, 3\%.}
\end{figure}

In Figure~\ref{FigLepJets} we compare the number of events predicted by our analysis to those of ATLAS 
as a function of $c\tau$ assuming 100\% Higgs branching.
The obtained $\approx 330$ events at $c\tau=4.7$~cm is an underprediction
compared to the full simulation results of $600\pm 40$,
most likely due to the stricter barrel/endcap regions employed.
For alternative lifetimes
we cannot come up with a physical explanation that could account for the shape discrepancy between
our curve and the reweighted result of ATLAS.

Now we wish to reinterpret the ATLAS analysis for $h\to\gamma_d\gamma_d$ 
decays predicted by the vector portal model described in Sec.~\ref{SecModels}.
A signal sample $gg\to h \to \gamma_d\gamma_d$ for $m_{\gamma_d}=400$~MeV 
was generated in \textsc{Pythia} and fed through our analysis.
Fig.~\ref{FigLepJets} shows the total number of events predicted as a function of $c\tau$,
as well as broken down by event Type.
More events are predicted than in the FRVZ model, and they peak at a lower $c\tau$,
since on average the $\gamma_d$ are more boosted.
The 95\% CL upper limit of $\approx 120$ ($\approx 30$) on the total (total excluding Type 2-2) number of events
can be inferred from the ATLAS plots. 
These numbers can be used along with Fig.~\ref{FigLepJets} to limit exotic Higgs branching fractions
for $m_{\gamma_d}=400$~MeV.

For alternative masses, since $E\gg m$, the properties of the simulated $\gamma_d$ will be approximately
the same but for the boost $\gamma=E/m_{\gamma_d}$.
If the efficiencies do not change significantly 
surrounding $m_{\gamma_d}=400$~MeV, which according to ATLAS is 
at least a good assumption for $0.25\lesssim m_{\gamma_d}/\text{GeV} \lesssim 1.5$ \cite{Aad2014yea},
then according to Eq.~\ref{EqPDecay} 
the number of events plotted as a function of $m_{\gamma_d}c\tau$ remains invariant.
Limits for alternative masses can be derived from the $m_{\gamma_d}=400$~MeV results using this observation.
In Fig.~\ref{FigLepResults} we present an example exclusion plot 
derived from Fig.~\ref{FigLepJets} in this way: the limit on
$Br(h\to \gamma_d\gamma_d)\times Br(\gamma_d\to \mu\mu)^2$ as a function of $m_{\gamma_d}$ and $c\tau$.
This is also a good approximation for the limit on $Br(h\to ss)\times Br(s\to \mu\mu)^2$.
In Sec.~\ref{SecPortalLims}, Fig.~\ref{FigLepResults} (and related limits on branchings to the other event Types)
are reinterpreted to bound mixing-mass parameter space for the Higgs and vector portals;
there the exclusion is extended up to the $\tau\tau$ threshold $m_{\gamma_d}\approx3.5$~GeV.

\begin{figure}[h]
 \centering
 \includegraphics[width=0.88\textwidth]{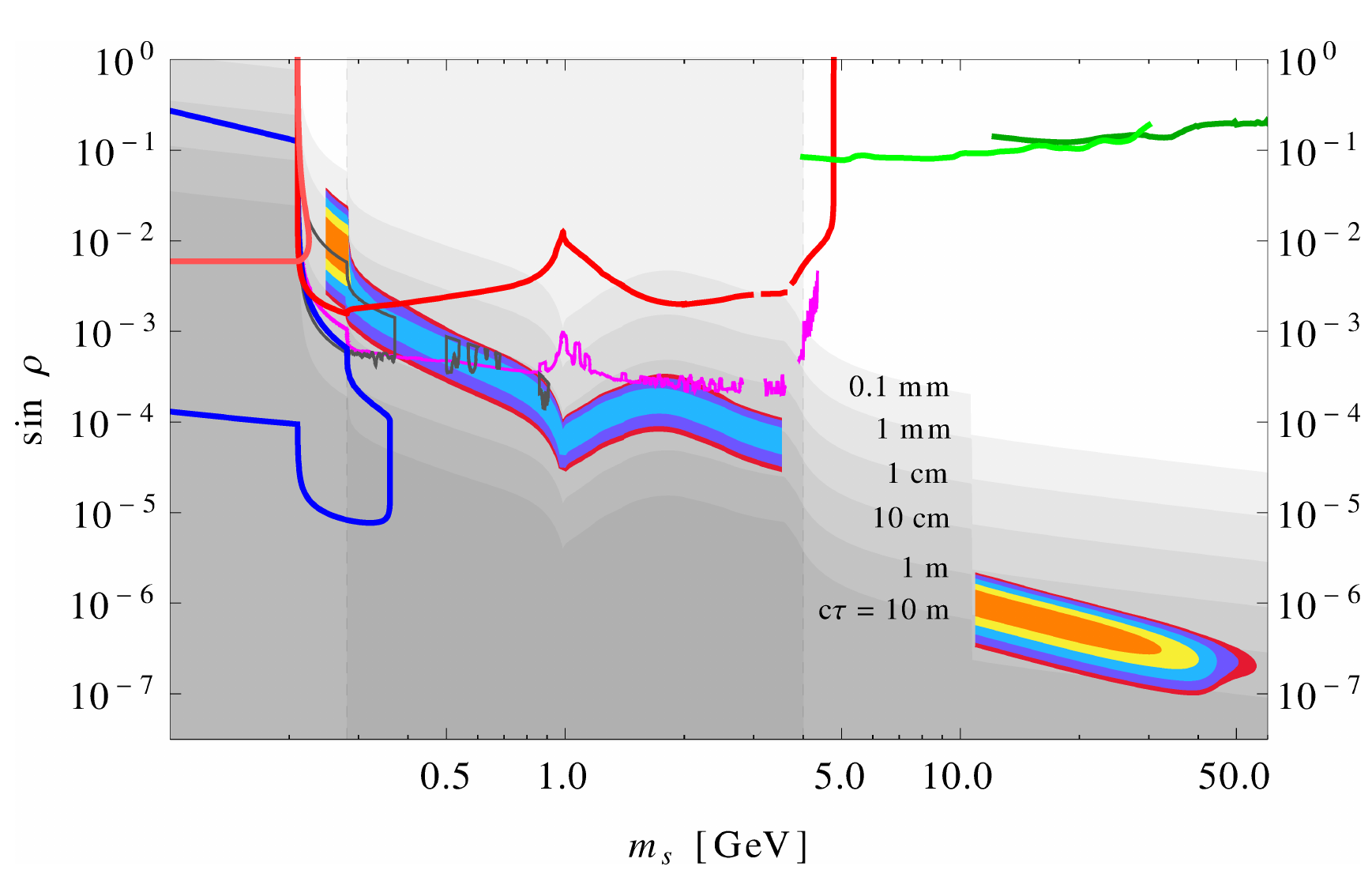}
 \caption{Exclusion plot for the real singlet scalar (Higgs) portal (see text).
 Incremental shadings mark areas of non-negligible lifetime.
 The coloured regions (this analysis) mark the exclusions assuming
 $Br(h\to ss)=$~30\%, 20\%, 10\%, 5\%, 3\%.
 }
 \label{FigParameterspaceHig}
\end{figure}

\begin{figure}[h]
 \centering
 \includegraphics[width=0.88\textwidth]{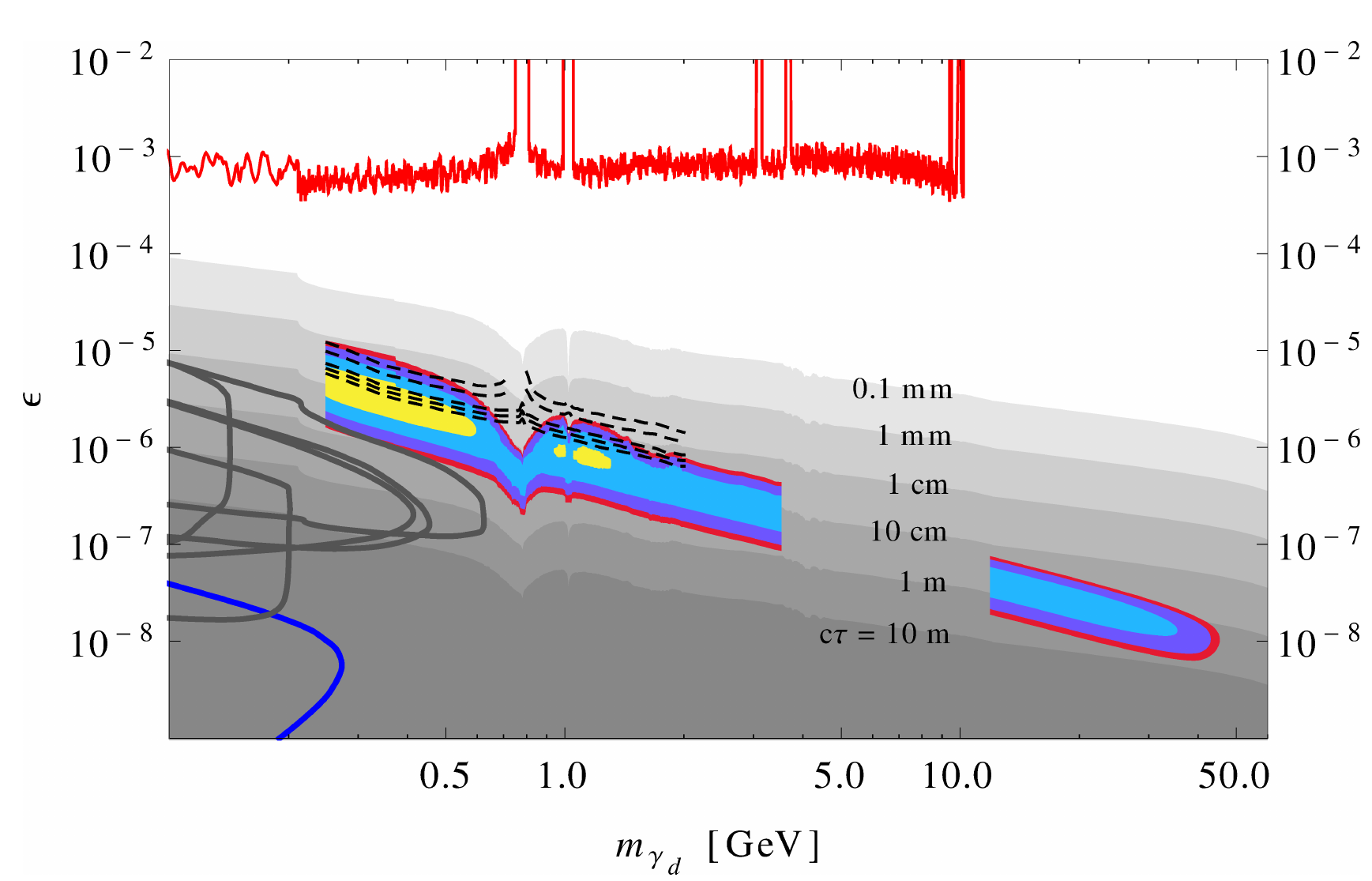}
 \caption{Exclusion plot for the vector portal (see text).
 Incremental shadings mark areas of non-negligible lifetime.
 The coloured regions (this analysis) and dashed lines (CMS Ref.~\cite{Khachatryan2015wka}) 
 mark the exclusions assuming
 $Br(h\to \gamma_d\gamma_d)=$~30\%, 20\%, 10\%, 5\%, 3\%.
 }
 \label{FigParameterspaceVec}
\end{figure}

\section{Portal limits \label{SecPortalLims}}

In Figs.~\ref{FigParameterspaceHig} and \ref{FigParameterspaceVec} 
we reinterpret the analyses of Sec.~\ref{SecResults} for the Higgs and vector portal models.
The coloured regions mark the exclusion assuming $Br(h\to ss)=$~30\%, 20\%, 10\%, 5\%, 3\%.

For the Higgs portal Fig.~\ref{FigParameterspaceHig},
we ask the reader to bear in mind that within the shaded region $2m_\pi<m_s\lesssim 4$~GeV
the branching fractions and lifetimes are known to be uncertain 
by up to an order of magnitude (see Ref.~\cite{Clarke2013aya}).
We adopt the most recent calculation \cite{Donoghue1990xh} below 1.4~GeV 
and smoothly interpolate to a perturbative calculation \cite{Clarke2013aya} above 2~GeV.
The limits that can be derived from Fig.~\ref{FigLepJets} (such as in Fig.~\ref{FigLepResults}) 
are independent of these uncertainties, so that there is enough information 
provided to reinterpret our results for alternative branching fractions and lifetimes;
in all cases there is excluded parameter space with $Br(h\to ss)<10\%$.
Limit lines (borrowed from Ref.~\cite{Clarke2013aya}) are from prompt $B$ decays at Belle/BaBar 
(reds) \cite{Chen2007zk,delAmoSanchez2010bk,Wei2009zv,Aubert2008ps}, 
the CHARM beam dump (blue) \cite{Bergsma1985qz,Bezrukov2009yw}, and LEP (greens) \cite{Buskulic1993gi,Acciarri1996um,Barate2003sz}.
Note that prompt searches at LHCb will be more sensistive for $1\lesssim m_s/\text{GeV} \lesssim 5$,
but non-trivial long-lifetime effects must be carefully taken into account \cite{SchmidtHoberg2013hba}.
Very recently, LHCb presented a search \cite{Aaij2015tna} 
for long-lived light scalars in $B^0\to K^{*0} s \to (K^\pm\pi^\mp)(\mu^+\mu^-)$ decays;
the limit shown in magenta is reinterpreted from their Figure 4 
assuming the branching expression found in Ref.~\cite{Batell2009jf}.
Also included are the excluded regions from a recent BaBar search for long-lived particles (dark grey) \cite{Lees2015nla},
reinterpreted from their Figure 3 (1~cm~$\le c\tau_s \le 100$~cm)
assuming $Br(B\to s X_s)\approx5.6\sin^2\rho\left(1-m_s^2/m_B^2\right)^2$ \cite{Grinstein1988yu}.

For the vector portal Fig.~\ref{FigParameterspaceVec},
we calculated lifetimes and branchings for $m_{\gamma_d}<12$~GeV using the 
$R(s)=\sigma(e^+e^-\to\text{hadrons})/\sigma(e^+e^-\to\mu^+\mu^-)$
values collated by the PDG \cite{Agashe2014kda}.
For $m_{\gamma_d}\ge 12$~GeV we used those values provided in Ref.~\cite{Curtin2014cca}.
The exclusion lines are from a BaBar search (red) \cite{Lees2014xha}, supernovae cooling (blue) \cite{Dent2012mx}, 
and various beam dump experiments (dark grey) \cite{Blumlein2011mv,Blumlein2013cua,Gninenko2012eq,Bjorken2009mm,Essig2010gu}.
Limits from a recent CMS search \cite{Khachatryan2015wka} 
for prompt $h\to \gamma_d\gamma_d$ decays are reproduced as dashed lines
assuming $Br(h\to \gamma_d\gamma_d)=$~30\%, 20\%, 10\%, 5\%, 3\%, bottom-to-top.

\section{\label{SecConclusion}Conclusion}

Searching for the displaced decays of long-lived neutral states is a sensitive way to look for exotic Higgs physics.
Already, searches at the LHC have probed branching fractions at the per cent level.

In Sec.~\ref{SecModels} we demonstrated how simple it is to build 
natural models with displaced Higgs decay phenomenology.
Those models taken together serve to emphasise two points:
(a) that long-lived neutral states can inherit a wide variety of couplings to SM particles, and
(b) in extended models, many possible production mechanisms and final states exist.
This motivates collaborations to present their results in as model-independent a way as is possible,
and in Sec.~\ref{SecResults} we sought to explore by example how this might be done.

Unlike for prompt events, no public tool exists to simulate the detector response to displaced decays.
Hence phenomenologists wanting to reinterpret displaced searches 
are reliant upon efficiencies provided by the collaborations.
In general, the reconstruction efficiency for a long-lived particle will depend on 
its mass $m$, transverse momentum $p_T$, lab-frame decay length $L$, pseudorapidity $\eta$, and its decay mode.
Ideally the phenomenologist would know the reconstruction efficiency 
as a function of all five of these parameters together,\footnote{Although 
it is only possible to provide a two-dimensional efficiency plot on paper,
we see no reason why collaborations couldn't provide higher dimensional plots as an online resource.}
and then the simple Monte Carlo method we have described in Sec.~\ref{SecResults},
requiring no simulation of displaced decays,
could be used to determine the reconstruction probability for any event.
In the interests of simplicity, the dependence on $\eta$ is likely to be weak enough to be split into barrel and endcap regions,
and then the reconstruction efficiencies could be provided as three-dimensional $(m,p_T,L)$
functions for each final state in the barrel/endcap.
In certain limits the dependence on one of these variables might even be removed.
For example, in the limit $E\gg m$, the efficiency dependence on $m$
for hadronic jets is expected to be weak.
The trigger efficiency, defined as the probability of triggering given reconstruction,
could subsequently be taken into account in a similar way.

Although this efficiency table wishlist was not fully realised
for either of the ATLAS displaced searches considered in Sec.~\ref{SecResults}, 
we were still able to demonstrate the principles of a simple Monte Carlo method for reinterpretation,
and we used it to constrain parameter space of interest for portal models (see also Sec.~\ref{SecPortalLims}).
Our hope is that this paper inspires the following take-home message regarding displaced searches:
if the relevant multidimensional efficiency tables are provided,
then phenomenologists will be able to reinterpret searches in the context of their own models.

\acknowledgments

The author would like to thank Raymond Volkas and Robert Foot for invaluable support and advice,
as well as James Barnard and Michael Schmidt for useful conversations.
This work was supported in part by the Australian Research Council.

\bibliographystyle{JHEP.bst}
\bibliography{references}

\end{document}